\documentstyle[12pt]{article}

\newcommand{\beq}[1]{\begin{equation} \label{#1} }
\newcommand{\eeq}   {\end{equation}}
\newcommand{\ds}{\displaystyle \mathstrut}

\newcommand{\Frac}[2]{\frac
{\textstyle\lefteqn{\phantom{{}_{\mathstrut}^{\mathstrut}}} #1}
{\textstyle\lefteqn{\phantom{{}_{\mathstrut}^{\mathstrut}}} #2}}

\newcommand{\vmu}{\mbox{\boldmath $\mu$}}

\newcommand{\av}[1]{\langle #1 \rangle}

\begin{document}

\title{Spin-orbit interaction in final state as possible reason for T-odd correlation in ternary fission}
\author{A.L.Barabanov\\
{\it Kurchatov Institute, Moscow 123182, Russia}}
\date{}
\maketitle

\begin{abstract}
A model for ternary fission is discussed in which a third particle ($\alpha$-particle) is emitted due to non-adiabatic change of the nuclear potential at neck rapture. An expression for energy and angular distribution of $\alpha$-particles is proposed.  It is shown that an interaction between spin of fissioning system and orbital momentum of $\alpha$-particle (spin-orbit interaction in the final state) results in recently observed asymmetry of $\alpha$-particle emission, which can be formally related to T-odd correlation. No strong dependence of the asymmetry on the angle of $\alpha$-particle emission with respect to the fission axis is predicted by the model in accordance with the experimental data.
\end{abstract}

\section{Introduction}

Recently T-odd correlation
\beq{1.1}
B=\left(\hat{\bf s}\left[\hat{\bf p}_{LF}\times\hat{\bf p}_{TP}\right]\right)
\eeq
was observed in ternary fission of $^{233}$U and $^{235}$U nuclei by cold polarized neutrons, where $\hat{\bf s}$ is the unit vector along the neutron spin ${\bf s}$, and $\hat{\bf p}_{LF}$ and
$\hat{\bf p}_{TP}$ are the unit vectors along the momenta ${\bf p}_{LF}$ and
${\bf p}_{TP}$ of light fragment and $\alpha$-particle (third particle), respectively \cite{1}-\cite{3}.

One used longitudinally polarized neutrons. Let us assume that the axis $y$ is aligned with the neutron beam, thus the vector $\hat{\bf s}$ is directed with or against the axis $y$. The target is placed at the origin of the coordinates. Fragment counters are aligned so that they fix fragments moving with and against the axis~$z$. Plates with $\alpha$-particle counters are disposed bilaterally along the target and transversely to the axis~$x$.

Let us assume that the frame of axes is right, the vector $\hat{\bf s}$ is aligned with the axis $y$, and the vector $\hat{\bf p}_{LF}$ is aligned with the axis $z$. Therefore, when $\alpha$-particle is emitted along the axis $x$, then $B=+1$, and $B=-1$ for the opposite direction of $\alpha$-particle emission. Denote the count rate of $\alpha$-particles for fixed $B$ by $N(B)$. Thus, the asymmetry
\beq{1.2}
D=\frac{N(+1)-N(-1)}{N(+1)+N(-1)}
\eeq
is measured. To exclude systematic errors one performs the measurements for different directions of the vectors $\hat{\bf s}$ and $\hat{\bf p}_{LF}$ with respect to the axes $y$ and $z$. The results of measurements are: $D=-(2.52\pm 0.14)\cdot 10^{-3}$ for $^{233}$U and
$D=+(0.83\pm 0.11)\cdot 10^{-3}$ for $^{235}$U.

The correlation (\ref{1.1}) is T-odd, i.e. it changes the sign along with the vectors $\hat{\bf s}$, $\hat{\bf p}_{LF}$ and $\hat{\bf p}_{TP}$ at time reversion. However, its observation cannot be considered as an evidence for violation of time-reversal invariance. Similar T-odd correlation
\beq{1.3}
\left(\hat{\bf s}\left[\hat{\bf p}_n\times\hat{\bf p}_{LF}\right]\right),
\eeq
where $\hat{\bf p}_n$ is the unit vector along the momentum of an incident neutron, is known for a long time in fission of nuclei by polarized slow neutrons (see, e.g. \cite{4}). It manifests itself as left-right asymmetry of light fragment emission with respect to the plane, formed by the vectors $\hat{\bf s}$ and $\hat{\bf p}_n$.

The same left-right asymmetry is observed in elastic scattering of any transversely polarized particle by nuclei. This asymmetry results from the correlation
\beq{1.4}
\left(\hat{\bf s}\left[\hat{\bf p}\times\hat{\bf p}'\,\right]\right),
\eeq
where $\hat{\bf s}$ is the unit vector along the spin of the incident particle, and $\hat{\bf p}$ and $\hat{\bf p}'$ are the unit vectors along the momenta of the incident and scattered particles, respectively. The reason for this correlation is a spin-orbit nuclear interaction $\sim {\bf s}{\bf l}$, where ${\bf l}$ is the relative orbital momentum of the incident particle and target nucleus.

Let us assume that T-odd correlation (\ref{1.1}) in ternary fission also results from spin-orbit interaction in the exit channel. This work is devoted to the theoretical analysis of this hypothesis. Note, that $\alpha$-particle emitted in ternary fission has zero spin. Thus, we mean the interaction
$\sim {\bf J}{\bf l}$ of spin ${\bf J}$ of the fissioning nucleus after $\alpha$-particle emission and relative orbital momentum ${\bf l}$ of the $\alpha$-particle and nucleus. Evidences for a spin-orbit interaction of such type between target nucleus spin and relative orbital momentum were presented, e.g. in \cite{5}.

\section{Naive estimation}

Let us start from a naive estimation of possible effect using analogies with classical mechanics and electrodynamics. As the result of a polarized neutron capture a fissioning nucleus is also polarized. Thus, the magnetic moment $\vmu$ of the nucleus is aligned with the axis $y$. Suppose, it generates a magnetic field of the dipole type outside the nucleus
\beq{2.1}
{\bf H}=\frac{3{\bf n}(\vmu{\bf n})-\vmu}{r^3},
\eeq
where ${\bf n}={\bf r}/r$ is the unit vector along the radius-vector. Therefore, the magnetic field is normal to the plane $(x,z)$, in which $\alpha$-particles move in the experiment \cite{1,2}. Let us assume for definiteness that the vector $\vmu$ is directed against the axis $y$, then everywhere in the plane $(x,z)$ outside the nucleus the vector ${\bf H}$ is directed along the axis $y$.

The $\alpha$-particle is affected by the Lorentz force
\beq{2.2}
{\bf F}=\frac{2e}{c}\left[{\bf v}\times{\bf H}\right],
\eeq
resulting its sideways deviation (here $e$ is the elementary charge, and ${\bf v}$ is the velocity of the $\alpha$-particle). When the $\alpha$-particle moves along the axis $x$, it deviates in a direction of the axis $z$, while the $\alpha$-particle moving in the opposite direction deviates against the axis $z$.

Consider the situation when a light fragment is emitted along the axis $z$. It is known that the maximum of angular distribution of $\alpha$-particle with respect to the momentum ${\bf p}_{LF}$ of the light fragment falls at the angle $\theta=82^0$ due to stronger Coulomb repulsion from the heavy fragment. Therefore, in this classical picture the magnetic field deviates the maximum of the angular distribution of the $\alpha$-particles emitted along the axis $x$ toward the edge of the plate, formed by $\alpha$-particle counters. At the same time for the $\alpha$-particles emitted against the axis $x$ the shift of the maximum of the angular distribution is directed toward the center of the plate of $\alpha$-particle counters. Thus, one can expect a difference in counting of $\alpha$-particles, emitted along and against the axis $x$, i.e. the observing asymmetry.

Let us give the quantitative estimate of the effect. The magnetic field goes down as $1/r^3$, therefore, it is naturally to assume that the effect is related to a small region of size $R$, where $R$ is the radius of the fissioning nucleus. We assume that in this region the velocity $v$ of $\alpha$-particle is constant. During the time $\Delta t=R/v$ the $\alpha$-particle is affected by a moment of the Lorentz force
\beq{2.3}
F\Delta t\simeq\frac{2eRH}{c}\simeq
\frac{2e\mu}{cR^2},
\eeq
which is equal to an increase of the transversal momentum
$\Delta p_{\perp}=m_{\alpha}\Delta v_{\perp}$ of the $\alpha$-particle (here $m_{\alpha}=4m$ is the mass of the $\alpha$-particle, where $m$ is the nucleon mass). It gives the angle of transversal deviation
\beq{2.4}
\Delta\theta=\frac{\Delta v_{\perp}}{v}=
\frac{e\mu}{2mcvR^2}.
\eeq

The asymmetry $D$ is of the scale of the dimensionless parameter $\Delta\theta$. Taking for $\mu$ the nuclear magneton $e\hbar/2mc$, we get
\beq{2.5}
D\sim\frac{e^2}{\hbar c}\,\frac{c}{v}
\left(\frac{\hbar}{mcR}\right)^2.
\eeq
The appearance of the fine structure constant $e^2/\hbar c=1/137$ is natural, because we consider here the electromagnetic binding between the charged $\alpha$-particle and the magnetic field of the fissioning nucleus.

In quantum mechanics this interaction is determined by the Hamiltonian
\beq{2.6}
\hat H=\frac{\left(\hat{\bf p}-e\hat{\bf A}/c\right)^2}{2m_{\alpha}},
\eeq
where $\hat{\bf A}$ is the operator of vector potential of the electromagnetic field. In the case being considered it is generated by the magnetic dipole and has the form
\beq{2.7}
\hat{\bf A}=\frac{\left[\hat{\vmu}\times{\bf r}\right]}{r^3},
\eeq
where $\hat{\vmu}$ is the operator of nuclear magnetic moment. Neglecting the term $\sim A^2$ we rewrite the Hamiltonian as follows
\beq{2.8}
\hat H=\frac{\hat{\bf p}^2}{2m_{\alpha}}-
\frac{e}{m_{\alpha}c}\,\frac{\hat{\vmu}\hat{\bf l}}{r^3},
\eeq
where $\hat{\bf l}=\left[{\bf r}\times\hat{\bf p}\right]$ is the operator of orbital momentum. Taking into account that the magnetic moment of the nucleus is aligned with its spin, i.e. $\hat{\vmu}=\mu\hat{\bf J}/J$, we find that the electromagnetic spin-orbit interaction
$\sim\hat{\bf J}\hat{\bf l}$ is responsible for the sideways deviation of the $\alpha$-particle in the magnetic field.

But in the region of size $R$ the nuclear spin-orbit interaction $\sim\hat{\bf J}\hat{\bf l}$ is at least two order of magnitude larger, because it does not include electromagnetic constant $e^2/\hbar c$. Thus, one can expect that the effect resulting from the nuclear spin-orbit interaction is of the scale
\beq{2.9}
D\sim\frac{c}{v}
\left(\frac{\hbar}{mcR}\right)^2.
\eeq
Taking as rough estimate $R\sim 10^2\,\hbar/mc\sim 10^{-12}$~cm and $v\sim 10^{-2}\,c$ we get the required magnitude $D\sim 10^{-2}$.

In a literal sense the picture proposed can be tested by studying the shifts of the maxima of the angular distributions of $\alpha$-particles emitted along and against the axis $x$ for given vectors ${\bf s}$ and ${\bf p}_{LF}$. These shifts would lead to a strong dependence of the counting rate asymmetry on the angle $\theta$ between the momenta ${\bf p}_{TP}$ and ${\bf p}_{LF}$. It was analyzed with the use of pairs of $\alpha$-particle counters corresponding to different angles $\theta$, however, no strong dependence was found (see \cite{1,2}). A general form for the probability $dw$ of $\alpha$-particle emission into the solid angle $d\Omega$ was proposed
\beq{2.10}
\frac{dw}{d\Omega}\sim
(1+DB)F(\cos\theta),
\eeq
being in accordance with experimental data. Indeed, let us fix the vectors ${\bf s}$ and
${\bf p}_{LF}$ along the axes $y$ and $z$, respectively. Thus, the probability of $\alpha$-particle emission in the direction fixed by polar $\theta$ and azimuthal $\varphi$ angles is given by
\beq{2.11}
\frac{dw}{d\Omega}\sim
(1+D\sin\theta\cos\varphi)F(\cos\theta).
\eeq
Here $\varphi$ is the angle between the projection of the vector ${\bf p}_{TP}$ on the $(x,y)$ plane and the axis $x$. In the experiment one fixes the $\alpha$-particles with $\theta\simeq\pi/2$, thus taking into account that $N(+1)\sim dw(\varphi\simeq 0)/d\Omega$ and $N(-1)\sim{}$ $dw(\varphi\simeq\pi)/d\Omega$ we obtain for the asymmetry (\ref{1.2}) the quantity $D$.

The classical and naive picture collapses when tested by experiment. However, it hardly means the failure of spin-orbit mechanism to produce T-odd correlation (\ref{1.1}) in ternary fission. Note, that there is no classical limit for spin of quantum particle, therefore, "magnetic field", related to spin of the fissioning nucleus, cannot be treated consistently as classical field. This is especially true for the nuclear field of spin-orbit forces which is certainly non-classical.

Thus, the problem is to construct the quantum model of ternary fission and to analyze in its frame a possible role of spin-orbit interaction in formation of T-odd correlation.

\section{Model for ternary fission}

The phenomenon of $\alpha$-decay is a manifestation of quantum tunneling of $\alpha$-particle through the barrier formed by attractive nuclear and repulsive Coulomb forces. Thus, one supposes that the wave function of the parent nucleus $^Z\!A$ includes with significant amplitude the component, corresponding to two interacting by nuclear and Coulomb forces clusters -- daughter nucleus $^{(Z-2)}(A-4)$ and $\alpha$-particle.

Evidently, it is the case for fissioning nucleus also. Before scission to two fragments $^{Z_L}\!A_L$ and $^{Z_H}\!A_H$ the wave function of the parent nucleus $^Z\!A$ ($Z_L+Z_H=Z$, $A_L+A_H=A$) is mainly represented by two-cluster ($^{Z_L}\!A_L+{}^{Z_H}\!A_H$) component. However, there is no reason for supression of a three-cluster component
$^{Z_L}\!A_L+{}^{Z_H}\!A_H+\alpha$ ($Z_L+Z_H+2=Z$, $A_L+A_H+4=A$) in the wave function of the parent nucleus $^Z\!A$.

As a result of a transience of fission process this three-cluster component cannot manifest itself via the tunneling through the barrier. But sharp changing of nuclear form and, consequently, of the nuclear potential can initiate non-adiabatic transition of $\alpha$-particle to continuum from bound or quasibound state. Let us assume that such a transition results in ternary fission.

The most sharp change of nuclear potential evidently takes place at neck rapture. Therefore, non-adiabatic transition populates continuum states described by wave functions enhanced near the neck. It corresponds to usual belief that in ternary fission $\alpha$-particle is emitted from the neck, supported by classical trajectory calculations.

Thus, we assume that $\alpha$-particle is emitted during the short time $\tau$ due to non-adiabatic change of nuclear potential at scission. At the final moment $t=\tau$ we have the $\alpha$-particle and two fragments $^{Z_L}\!A_L$ and $^{Z_H}\!A_H$ with a distance $R_f$ between their center-of-masses. The origin of coordinates is in the center-of-mass of two fragments. The centers of fragments are located at the axis $z$, and this axis is directed from the heavy fragment to the light one. Potential energy of the $\alpha$-particle is given by the sum of potentials from both fragments
\beq{3.1}
U_f({\bf r})=U_L({\bf r}-\frac{A_H}{A_L+A_H}R_f\hat{\bf z})+
U_H({\bf r}+\frac{A_L}{A_L+A_H}R_f\hat{\bf z}),
\eeq
where $\hat{\bf z}$ is the unit vector along the axis $z$.

Really further acceleration of $\alpha$-particle occurs in the time dependent Coulomb field of two fragments removing one from the other. To simplify the model we neglect the fragment movement. Then, the wave function $\psi_{{\bf k}}^{(-)}({\bf r})$ of the $\alpha$-particle in the final state corresponding to the energy $E_f=\hbar^2k^2/2m_{\alpha}$ is given by a solution of the stationary Shroedinger equation with the potential (\ref{3.1}) with the asymptote
\beq{3.2}
\psi_{{\bf k}}^{(-)}({\bf r})
\enskip\mathop{\longrightarrow}\limits_{r\to\infty}\enskip
f(\hat{\bf k})\,\frac{e^{-ikr}}{r}+e^{i{\bf k}{\bf r}},
\eeq
where ${\bf k}$ is the wave vector of the $\alpha$-particle at infinity. Here and below we assume for simplicity that the fragment Coulomb fields are screened at large distances, i.e. go down faster than $1/r$.

Then let us assume that at the initial moment $t=0$ the potential energy of $\alpha$-particle is given by the sum of (\ref{3.1}) and the potential $U({\bf r})$ caused by the neck between the fragments. We denote $\psi_i({\bf r})$ a solution of the stationary Shroedinger equation with the potential
\beq{3.3}
U_i({\bf r})=U_f({\bf r})+U({\bf r})
\eeq
for the $\alpha$-particle with the energy $E_i$.

During the time $\tau$ the Hamiltonian changes
\beq{3.4}
\hat H_i=\frac{\hat{\bf p}^2}{2m_{\alpha}}+U_i({\bf r})
\quad\longrightarrow\quad
\hat H_f=\frac{\hat{\bf p}^2}{2m_{\alpha}}+U_f({\bf r}).
\eeq
Let us introduce the time dependent Hamiltonian
\beq{3.5}
\hat H(t)=\hat H_f+\hat V(t).
\eeq
The time dependent part has the limiting values
\beq{3.6}
\hat V(0)=U({\bf r}),\qquad
\hat V(\tau)=0,
\eeq
and we consider it as perturbation. Then in the first order of perturbation theory the probability of non-adiabatic transition from the initial state $\psi_i({\bf r})$ to the final continuum state $\psi_{{\bf k}}({\bf r})$ is given by (see \cite{6})
\beq{3.7}
w(i\to{\bf k})=
\frac{1}{(E_f-E_i)^2}
\left|\,\int\limits_0^{\tau}
\av{\psi_{{\bf k}}^{(-)}|\frac{d\hat V}{dt}|\,\psi_i}\,
e^{i\omega_{fi}t}\,dt\,\right|^2,
\eeq
where $\omega_{fi}=(E_f-E_i)/\hbar$.

In the experiment one fixes the $\alpha$-particles emitted into the solid angle $d\Omega$ and falling to the energy interval $dE_f$. Taking the normalization volume equal unity we sum (\ref{3.7}) over the final states, corresponding to the intervals $d\Omega$ and $dE_f$. Thus, we obtain for the differential probability of $\alpha$-particle emission into continuum
\beq{3.8}
\frac{dw}{d\Omega dE_f}=
\frac{m_{\alpha}\sqrt{2m_{\alpha}E_f}}
{(2\pi\hbar)^3(E_f-E_i)^2}
\left|\,\int\limits_0^{\tau}
\av{\psi_{{\bf k}}^{(-)}|\frac{d\hat V}{dt}|\,\psi_i}\,
e^{i\omega_{fi}t}\,dt\,\right|^2.
\eeq
This formula determines angular and energy distribution of $\alpha$-particles in ternary fission in the given model. Integrating (\ref{3.8}) over all solid angles and energies we get the probability of ternary fission with respect to the binary one.

\section{Scheme of calculations}

To calculate the total and differential probabilities of ternary fission given by (\ref{3.8}) we need the initial and final wave functions of $\alpha$-particle as well as the law $\hat V(t)$ determined by the dynamics of the final stage of fission process.

The potentials $U_i({\bf r})$ and $U_f({\bf r})$ acting to $\alpha$-particle in the initial and final states have azimuthal but not spherical symmetry. Thus the projection $m$ of the orbital momentum~$l$ of $\alpha$-particle on the axis $z$ conserves, but not the orbital momentum itself.

An eigenfunction $\psi_i({\bf r})$ of the initial Hamiltonian describing a bound or quasibound state corresponds to the definite energy $E_i$ and projection $m_i$. To be more precise, there are two degenerated states corresponding the energy $E_i$ and projections $m_i$ and $-m_i$, respectively. We take $\psi_i({\bf r})$ as a superposition
\beq{4.1.1}
\psi_i({\bf r})=\psi_{m_i}({\bf r})+e^{i\eta}\,\psi_{-m_i}({\bf r}),
\eeq
normalized to unity
\beq{4.1.2}
\int |\psi_i({\bf r})|^2d^3r=1.
\eeq
Here $\eta$ is a random phase. The function $\psi_{m_i}({\bf r})$ can be represented as the superposition over orbital momenta $l_i\ge m_i$
\beq{4.1}
\psi_{m_i}({\bf r})=\sum_{l_i}\,\av{\hat{\bf r}|l_im_i}\,f_{l_im_i}(r),\qquad
\av{\hat{\bf r}|l_im_i}=i^{l_i}Y_{l_im_i}(\hat{\bf r}).
\eeq
Putting this function to the Shroedinger equation
\beq{4.2}
\hat H_i\psi_i({\bf r})=E_i\psi_i({\bf r}),
\eeq
we obtain the coupled equations for radial functions
\beq{4.3}
\frac{d^2f_{l_im_i}}{dr^2}+
\frac{2}{r}\frac{df_{l_im_i}}{dr}-
\frac{l_i(l_i+1)}{r^2}f_{l_im_i}+
\frac{2m_{\alpha}E_i}{\hbar^2}f_{l_im_i}-
\frac{2m_{\alpha}}{\hbar^2}
\sum_{l'_i}\av{l_im_i|U_i|l'_im_i}f_{l'_im_i}=0.
\eeq
Together with the boundary conditions
\beq{4.4}
f_{l_im_i}(r)
\enskip\mathop{\longrightarrow}\limits_{r\to\infty}\enskip 0
\eeq
they determine the initial wave function.

Then we are looking for the wave function of the final state as a series in spherical harmonics
\beq{4.5}
\psi_{{\bf k}}^{(-)}({\bf r})=
\sum_{lm}\,\av{\hat{\bf r}|lm}\,R_{lm}^{(-)}(r).
\eeq
Substituting it into the Shroedinger equation
\beq{4.6}
\hat H_f\psi_{{\bf k}}^{(-)}({\bf r})=
E_f\psi_{{\bf k}}^{(-)}({\bf r})
\eeq
we get the coupled equations for the functions $R_{lm}^{(-)}(r)$ of the same type as (\ref{4.3}). Note that matrix elements of potential entering the coupled equations are diagonal on the projection $m$ due to azimuthal symmetry.

To satisfy the boundary condition (\ref{3.2}) it is convenient to take the radial functions in the form
\beq{4.7}
R_{lm}^{(-)}(r)=\sum_{l_0}4\pi Y^*_{l_0m}(\hat{\bf k})
\frac{F^m_{l\,l_0}(r)}{r}.
\eeq
We get for the functions $F^m_{l\,l_0}(r)$ corresponding to the projection $m$ of orbital momentum on the axis $z$
\beq{4.8}
\frac{d^2F^m_{l\,l_0}(r)}{dr^2}-
\frac{l(l+1)}{r^2}F^m_{l\,l_0}(r)+
k^2F^m_{l\,l_0}(r)-
\frac{2m_{\alpha}}{\hbar^2}
\sum_{l'}\av{lm|U_f({\bf r})|l'm}
F^m_{l'\,l_0}(r)=0,
\eeq
with boundary conditions
\beq{4.9}
F^m_{l\,l_0}(r)
\enskip\mathop{\longrightarrow}\limits_{r\to\infty}\enskip
\frac{1}{2k}\left((kr)h_l^{(+)}(kr)\delta_{l\,l_0}+
(kr)h_l^{(-)}(kr)S_m(l\to l_0)\right).
\eeq
The equations (\ref{4.8}) and (\ref{4.9}) describe a scattering of $\alpha$-particle by the deformed potential $U_f({\bf r})$ provided the ingoing state is a superposition over orbital momentum while the outgoing one is determined by asymptotically fixed orbital momentum $l_0$.

Substituting (\ref{4.1}),(\ref{4.5}) and (\ref{4.7}) into (\ref{3.8}) and averaging over random phase $\eta$ we get the double differential probability of $\alpha$-particle emission as series in Legandre polynomials
\beq{4.10}
\frac{dw}{d\Omega dE_f}=
\sum_{Q=0,1,2\ldots}(2Q+1)\,a_Q(E_f)\,P_Q(cos\theta).
\eeq
The energy dependent coefficients $a_Q(E_f)$ are determined by the formula
\beq{4.11}
a_Q(E_f)=\frac{m_{\alpha}\sqrt{2m_{\alpha}E_f}}{\pi^2\hbar^3(E_f-E_i)^2}
\sum_{l_0l'_0}\sqrt{\frac{2l_0+1}{2l'_0+1}}
C^{l'_00}_{l_00Q0}C^{l'_0m_i}_{l_0m_iQ0}
A(l_0m_i,E_f)A^*(l'_0m_i,E_f),
\eeq
where
\beq{4.12}
A(l_0m_i,E_f)=\sum_{l\,l_i}
\int\limits_0^{\tau}
\av{\frac{F^{m_i}_{l\,l_0}(r)}{r}i^lY_{lm_i}(\hat{\bf r})|
\frac{d\hat V}{dt}|f_{l_im_i}(r)i^{l_i}Y_{l_im_i}(\hat{\bf r})}
e^{i\omega_{fi}t}dt.
\eeq
We take into account azimuthal symmetry of the perturbation $\hat V$.

The total probability of ternary fission with respect to the binary one is given by integral over solid angle and energy of isotropic term in (\ref{4.10})
\beq{4.13}
w=4\pi\int\limits_0^{\infty}a_0(E_f)dE_f,
\eeq
where
\beq{4.14}
a_0(E_f)=\frac{m_{\alpha}\sqrt{2m_{\alpha}E_f}}
{\pi^2\hbar^3(E_f-E_i)^2}
\sum_{l_0}\left|A(l_0m_i,E_f)\right|^2.
\eeq

\section{T-odd correlation}

As the result of slow neutron capture by a target nucleus with spin $I$ a compound nucleus arises with spin $J_c$, where $J_c=I-1/2$ or $J_c=I+1/2$. If  the neutron polarization is equal to $p_n$ and the target nucleus is not oriented, then the polarization of compound nucleus with spin $J_c\ne 0$ is
\beq{5.1}
p(J_c)=\left\{\begin{array}{ll}
-\,\Frac{1}{3}\,p_n, & J_c=I-\Frac{1}{2},
\\[\bigskipamount]
\Frac{2I+3}{3(2I+1)}\,p_n, & J_c=I+\Frac{1}{2}.
\end{array}\right.
\eeq
The polarization axis for compound nucleus coincides with that one for captured neutron. Below as elsewhere above we take the axis $y$ along the neutron beam as the polarization axis.

In a binary fission the spin ${\bf J}_c$ transforms to the sum of fragment spins
${\bf J}_L$ and ${\bf J}_H$ and a relative orbital momentum ${\bf L}$ between two fragments. Let us assume that in a ternary fission the nucleus before scission emits an $\alpha$-particle with a small initial orbital momentum ${\bf l}$, therefore the residual spin ${\bf J}$ can be taken approximately equal to ${\bf J}_c$ and $p(J)\simeq p(J_c)$.

Thus, there is a fissioning nucleus with spin ${\bf J}$ aligned with the axis $y$. If its state with a projection $M$ of spin $J$ on the axis $z$ is described by the function $\Psi_{JM}$, then generally a pure quantum state of the nucleus is given by superposition
\beq{5.2}
\Psi_J=\sum_Ma_M(J)\Psi_{JM}.
\eeq
A spin state of an ensemble of such nuclei is given by density matrix, averaged over the ensemble, namely,
\beq{5.3}
\rho_{MM'}(J)=\av{a_M(J)a^*_{M'}(J)}.
\eeq
In the case being considered the density matrix takes the form
\beq{5.4}
\rho_{MM'}(J)=\frac{1}{2J+1}\left(\delta_{MM'}+
3\,p(J)
\sqrt{\frac{J}{J+1}}\,\sqrt{\frac{4\pi}{3}}
\sum_qC^{JM'}_{JM1q}Y_{1q}(\hat{\bf y})\right).
\eeq

Now let us assume that there is a nuclear interaction depending on a relative direction of the spin ${\bf J}$ and the orbital momentum ${\bf l}$ of an $\alpha$-particle. Since
${\bf J}={\bf J}_L+{\bf J}_H+{\bf L}$ this spin-orbit interaction $\sim {\bf J}{\bf l}$ may be an effective manifestation of interactions really arising between the angular momenta ${\bf J}_L$ and ${\bf l}$, ${\bf J}_H$ and ${\bf l}$ or ${\bf L}$ and ${\bf l}$. Omitting the discussion about the nature of the spin-orbit interaction we include it as a small correction into the perturbation
\beq{5.5}
\hat V=\hat V_0+\left(\hat V_{Jl}(\hat{\bf J}\hat{\bf l})+
(\hat{\bf J}\hat{\bf l})\hat V_{Jl}\right).
\eeq
Here the spin-orbit term is written in symmetric, evidently Hermitian form. This term results in changing of the spin wave function of fissioning nucleus in the process of $\alpha$-particle emission. Therefore the model for ternary fission described above should be generalize to take into account the angular momenta.

To do this note that the differential probability of ternary fission caused by perturbation (\ref{5.5}) is given by the equation (\ref{3.8}), where the following replacements should be made
\beq{5.6}
\begin{array}{lll}
\psi_i({\bf r}) & \longrightarrow &
\psi_i({\bf r})\Psi_J=
\psi_i({\bf r})\sum\limits_Ma_M(J)\Psi_{JM},
\\[\bigskipamount]
\psi_{{\bf k}}^{(-)}({\bf r}) & \longrightarrow &
\psi_{{\bf k}}^{(-)}({\bf r})\Psi_{JM_f}.
\end{array}
\eeq
We obtain the probability of $\alpha$-particle emission to the intervals $d\Omega$ and $dE_f$ provided the nuclear system remains in the state with the projection $M_f$ of spin $J$ on the axis $z$. Since only $\alpha$-particle is registered, summation over $M_f$ should be performed.

It gives in linear in $\hat V_{Jl}$ approximation
\beq{5.7}
\frac{dw}{d\Omega dE_f}=
\sum_{Q=0,1,2\ldots}(2Q+1)a_Q(E_f)P_Q(\cos\theta)+
p(J)\cos\varphi\sum_{Q=1,2\ldots}(2Q+1)b_Q(E_f)
P^1_Q(\cos\theta),
\eeq
where
\beq{5.8}
P^1_Q(\cos\theta)=\sin\theta\,\frac{dP_Q(\cos\theta)}{d\cos\theta}
\eeq
is the reduced Legandre polynomial. The coefficient $b_Q(E_f)$ is of the form
\beq{5.9}
\begin{array}{l}
b_Q(E_f)=-\,\Frac{2m_{\alpha}\sqrt{m_{\alpha}E_f}}{\pi^2\hbar^3(E_f-E_i)^2}\,
\Frac{J}{\sqrt{Q(Q+1)}}\,
{\ds\sum_{l_0l'_0}}\,\sqrt{\Frac{2l_0+1}{2l'_0+1}}\,
C^{l'_00}_{l_00Q0}\times{}
\\[\bigskipamount]
\phantom{b_Q(E_f)}\times
{\rm Im}\left(A(l_0m_i,E_f)
\left(C^{l'_0m_i+1}_{l_0m_iQ\,1}
B^*(l'_0m_i\,1,E_f)-
C^{l'_0m_i-1}_{l_0m_iQ\,-\!1}
B^*(l'_0m_i\,-\!1,E_f)\right)\right),
\end{array}
\eeq
where
\beq{5.11}
\begin{array}{l}
B(l_0m_i\lambda,E_f)={}
\\[\bigskipamount]
\phantom{B}={\ds\sum_{l\,l_i}}\left(\sqrt{l_i(l_i+1)}C^{l_im_i+\lambda}_{l_im_i1\lambda}
{\ds\int\limits_0^{\tau}}
\av{\Frac{F^{m_i+\lambda}_{ll_0}}{r}\,
i^lY_{lm_i+\lambda}|
\Frac{d\hat V_{Jl}}{dt}|f_{l_im_i}i^{l_i}Y_{l_im_i+\lambda}}
e^{i\omega_{fi}t}dt+{}\right.
\\[\bigskipamount]
\left.\phantom{B={}}+
\sqrt{l(l+1)}C^{l\,m_i+\lambda}_{l\,m_i1\lambda}
{\ds\int\limits_0^{\tau}}
\av{\Frac{F^{m_i+\lambda}_{ll_0}}{r}\,
i^lY_{lm_i}|
\Frac{d\hat V_{Jl}}{dt}|f_{l_im_i}i^{l_i}Y_{l_im_i}}
e^{i\omega_{fi}t}dt\right).
\end{array}
\eeq
Note that the expression (\ref{5.7}) includes the needed asymmetry $\sim\cos\varphi$ of $\alpha$-particle emission along and against the axis $x$.

In the experiment \cite{1,2} one fixed $\alpha$-particles with $\theta\simeq\pi/2$ and
$\varphi\simeq 0$ or $\pi$. Thus, integrating (\ref{5.7}) over $E_f$, we get for the angular distribution the following general form
\beq{5.12}
\frac{dw}{d\Omega}=F_1(\cos\theta)+p(J)\,\tilde D\cos\varphi\, F_2(\cos\theta),
\eeq
where $\tilde D$ is a constant, what is close to (\ref{2.11}).

The asymmetry of $\alpha$-particle emission at the angles $\varphi=0$ and $\varphi=\pi$ is equal to
\beq{5.13}
\frac{N(+1)-N(-1)}{N(+1)+N(-1)}=p(J)\tilde D\,
\frac{F_2(\cos\theta)}{F_1(\cos\theta)}.
\eeq
It is natural to expect that both functions $F_1(\cos\theta)$ and $F_2(\cos\theta)$ formed by summation of a lot of terms, as well as their ratio, change smoothly as functions of $\theta$. Thus, it agrees qualitatively with the absence of strong dependence of the effect on the angle $\theta$ \cite{1,2}. Therefore, this experimental result cannot be considered as unambiguous evidence against spin-orbital mechanism of T-odd correlation formation.

Note that the function $F_2(\cos\theta)$ is given by summation of the terms including reduced Legandre polynomials (\ref{5.8}) which are equal to zero at $\theta=\pi/2$ for even indexes $Q=2,4\ldots$. It means that the effect is due to the odd in $Q$ terms arising as the result of mixing of states with even and odd orbital momenta (or even and odd parity) by the coupled equations (\ref{4.8}). This mixing is in its turn caused by the absence of symmetry of Hamiltonian or, to be more precise, of potential (\ref{3.1}) under space inversion because the direction of the axis $z$ is fixed from the heavy fragment to the light one. Thus, in the quantum model being considered T-odd correlation is directly related to the asymmetry between heavy and light fragments as well as in the naive classical approach described above.

\section{Conclusion}

The model for ternary fission is discussed in which a third particle ($\alpha$-particle) is emitted due to non-adiabatic change of the nuclear potential at neck rapture. An expression for energy and angular distribution of $\alpha$-particles is proposed. Neglecting the fragment motion during $\alpha$-particle acceleration strongly facilitates the model and does not allow to expect faithful reproducing of corresponding experimental data. However, one may hope to describe qualitatively the specific angular anisotropy of $\alpha$-particles in ternary fission.

It is shown that inclusion to the model of an interaction between spin of fissioning system and orbital momentum of $\alpha$-particle (spin-orbit interaction in the final state) leads to the asymmetry of $\alpha$-particle emission of the same type that the observed asymmetry related to T-odd correlation (\ref{1.1}). No strong dependence of the asymmetry on the angle of $\alpha$-particle emission with respect to the fission axis is predicted by the model in accordance with the experimental data.
\bigskip

I am grateful for helpful discussions to V.E.Bunakov, W.I.Furman, F.Goennenwein, G.A.Petrov and G.V.Danilyan. The work is supported by RFBR grant 00-15-96590 and INTAS grant 99-0229.

\end{document}